\documentclass[reprint]{revtex4-1}

\usepackage{amsmath}    
\usepackage{graphicx}   
\usepackage{verbatim}   
\usepackage{color}      
\usepackage{subfigure}  
\usepackage{hyperref}   

\begin{document}

\title{GaN/AlGaN 2DEGs in the quantum regime: Magneto-transport and photoluminescence to 60 tesla}

\author{S.~A. Crooker$^{1}$, M. Lee$^2$, R.~D. McDonald$^1$, J. L. Doorn$^{1}$, I. Zimmermann$^{1}$, Y. Lai$^{1}$, L. E. Winter$^{1}$, Y. Ren$^3$, Y.-J. Cho$^{2}$, B. J. Ramshaw$^4$, H. G. Xing$^{2,3}$, D. Jena$^{2,3}$}
\affiliation{$^1$National High Magnetic Field Laboratory, Los Alamos National Laboratory, Los Alamos, NM 87545}
\affiliation{$^2$School of Electrical and Computer Engineering, Cornell University, Ithaca, NY 14853}
\affiliation{$^3$Department of Materials Science and Engineering, Cornell University, Ithaca, NY 14853}
\affiliation{$^4$Laboratory of Atomic and Solid State Physics, Cornell University, Ithaca, New York 14853}

\date{\today}

\begin{abstract}
Using high magnetic fields up to 60~T, we report magneto-transport and photoluminescence (PL) studies of a two-dimensional electron gas (2DEG) in a GaN/AlGaN heterojunction grown by molecular-beam epitaxy. Transport measurements demonstrate that the quantum limit can be exceeded (Landau level filling factor $\nu < 1$), and show evidence for the $\nu =2/3$ fractional quantum Hall state. Simultaneous optical and transport measurements reveal synchronous quantum oscillations of both the PL intensity and longitudinal resistivity in the integer quantum Hall regime.  PL spectra directly reveal the dispersion of occupied Landau levels in the 2DEG and therefore the electron mass. These results demonstrate the utility of high (pulsed) magnetic fields for detailed measurements of quantum phenomena in high-density 2DEGs. 
\end{abstract}

\maketitle

The wide-bandgap semiconductor GaN is a foundational material for solid-state lighting applications and high-power electronics. Furthermore, the two-dimensional electron gas (2DEG) that forms naturally at GaN/AlGaN heterointerfaces \cite{Ambacher1999, Frayssinet2000, Elhamri2000} is of considerable interest for high-electron mobility transistors.  2DEG structures grown by molecular-beam epitaxy (MBE) have exhibited low-temperature electron mobilities exceeding $10^5$~cm$^2$/Vs \cite{Manfra2004, Skier2005}, galvanizing interest in quantum phenomena and novel electron correlations in GaN-based materials. Indeed, transport measurements have shown a robust integer quantum Hall effect (IQHE) in GaN/AlGaN heterojunctions \cite{Frayssinet2000, Elhamri2000, Manfra2004, Skier2005, Knap2004, Suzuki2018, Schmult2019, Kruckeberg2020}, and an indication of a fractional quantum Hall state (Landau level filling factor $\nu$=5/3) was reported by Manfra \textit{et al.} nearly two decades ago \cite{Manfra2002}.  In comparison with the more widely studied GaAs-based 2DEGs, electrons in GaN-based 2DEGs have significantly heavier effective masses ($\approx$0.24~$m_0$ versus $\approx$0.07~$m_0$ in GaAs, where $m_0$ is the bare electron mass), and the dielectric constant is smaller ($\epsilon$$\approx$9.5 in GaN versus $\approx$13 in GaAs), so that enhanced electron-electron interactions are expected. In this regard, 2DEGs in GaN more closely resemble those found in other wide-bandgap semiconductors such as ZnO, where significant progress has recently been made \cite{Falson2018}.

However, peak mobilities in GaN-based 2DEGs are, to date, typically achieved at relatively large electron densities $n_e \sim 10^{12}$/cm$^2$ \cite{ManfraMobility, Cho2019}, so that high magnetic fields $B \gtrsim$ 40~T are required to reach the so-called ``quantum limit'' wherein all the electrons reside in the lowest spin-polarized Landau level (\textit{i.e.}, $\nu \leq 1$). Such large $B$ are (just) within reach of modern superconducting-resistive hybrid magnet technologies, but are routinely exceeded by pulsed magnets \cite{Battesti2018}. Pulsed fields can therefore enable detailed studies of high-density 2DEGs, including not only transport but also optical measurements that probe the response of the 2DEG to a photogenerated hole, which have historically proven to be a very powerful tool to measure screening and many-body effects in GaAs- and ZnO-based systems \cite{Goldberg1992, Nurmikko1993, Kukushkin1996, Heiman1988, Katayama1989, Goldberg1990, Buhmann1990, Turberfield1990, Byszewski2006, Solovyev2015}.  

To this end, we report both transport and optical studies of a high-mobility 2DEG in a GaN/AlGaN heterojunction in pulsed magnetic fields to 60~T. We demonstrate that beyond the quantum limit, transport measurements show clear evidence for the $\nu =2/3$ fractional quantum Hall effect (FQHE) state. Moreover, simultaneous optical and transport studies reveal nearly-synchronous quantum oscillations of both the photoluminescence (PL) intensity and the longitudinal resistivity; however, the optical illumination required to perform PL significantly (and persistently) increases $n_e$ to the point where only $\nu \geq 3$ can be reached in 60~T in the same heterostructure.
 
The GaN/AlGaN structure (see inset, Fig. 1a) was grown by MBE on a semi-insulating single-crystal GaN substrate with low dislocation density ($\sim 5 \times 10^4$/cm$^2$), following Ref. \cite{Cho2019}.  After the initial growth of a 300~nm GaN buffer layer, a thin 21~nm Al$_{0.07}$Ga$_{0.93}$N barrier layer was grown. A high-mobility 2DEG formed naturally at the interface due to the spontaneous polarization discontinuity across the junction \cite{Ambacher1999}. The structure was capped by a final 3~nm GaN layer.  For transport studies, Ti/Au contacts were deposited and annealed at the corners of 3~mm $\times$ 3~mm squares in a van der Pauw geometry. The sample was mounted in a $^3$He cryostat in a 60~T capacitor-driven pulsed magnet. The magnet pulse pulse has a 9~ms rise time and total duration of $\approx$90~ms. Resistivity was measured using dc current, which avoids measurement-phase issues associated with high-frequency ac lock-in detection of high-resistance samples.  Appropriate combinations of current and magnetic field direction were used to symmetrize the data and accurately measure both the longitudinal ($R_{xx}$) and transverse ($R_{xy}$) magnetoresistance. 

Separately, photoluminescence (PL) measurements up to 60~T were performed on the same structure, which was mounted on a fiber-coupled probe and immersed in superfluid $^4$He at 1.5~K for optimal heat sinking. Unpolarized excitation light from the 325~nm (3.82~eV) line of a HeCd laser was directed to the sample, and PL was collected from the sample, using a multimode optical fiber. The PL polarization was not resolved. The PL was dispersed in a 300~mm spectrometer and measured by a fast charge-coupled device (CCD).  Full spectra were continuously acquired every 0.6~ms throughout the pulse \cite{Goryca2019}.

In a final set of measurements, resistivity and PL were simultaneously measured in a 18~T superconducting magnet, using a fiber-coupled transport probe with the sample in superfluid $^4$He at 1.5~K. AC current excitation at 17~Hz and lock-in detection provided a measure of the longitudinal resistance, while high-resolution PL spectra were detected using a 500~mm spectrometer and a liquid nitrogen cooled CCD. 

Figure 1a shows the transverse resistance $R_{xy}$ up to 60~T. Data acquired during both the upsweep and downsweep of the pulsed field are presented, to show the overall signal-to-noise, drifts, and data quality that are achievable using dc transport methods to measure 2D electron systems in pulsed magnets.  During these studies, it was observed that the 2DEG's conductivity varied from cooldown to cooldown, suggesting a history-dependent carrier density $n_e$. As demonstrated in early work \cite{Elhamri2000}, these variations arise from the history of optical illumination on the sample, as shown below. 

We focus first on the black curve in Fig. 1a, which was acquired after the sample had been three days in the dark at 300~K. Well-defined IQHE plateaus were observed, with the quantum limit ($\nu =1$) achieved at $B \simeq 27$~T, indicating 2DEG electron density $n_e =\nu e B/h \simeq 6.5 \times 10^{11}$/cm$^2$.  We note that while the IQHE has been reported many times in GaN-based 2DEGs \cite{Manfra2004, Skier2005, Knap2004, Suzuki2018, Schmult2019, Kruckeberg2020}, these measurements are the first to explore transport in the FQHE regime beyond the quantum limit. Crucially, $R_{xy}$ also shows an \textit{additional} plateau forming at approximately 41~T, which coincides with the expected position (and quantized resistance $R_{xy} = 3 h / 2 e^2$) of the $\nu = 2/3$ FQHE state. 

The longitudinal resistance $R_{xx}$ is shown in Fig. 1b, which was acquired during a separate cooldown, again after several days in the dark. Zero-resistance minima confirm the IQHE and a slightly larger $n_e$ ($\nu=1$ occurs at 29~T; thus $n_e \simeq 7.0 \times 10^{11}$/cm$^2$).  Most importantly, the pronounced dip at 44~T again strongly supports an interpretation in terms of the $\nu = 2/3$ FQHE state. While evidence for the fractional $\nu=5/3$ state in a GaN/AlGaN 2DEG was observed previously \cite{Manfra2002}, we emphasize that Fig. 1 demonstrates for the first time that phenomena beyond the quantum limit ($\nu<1$) are indeed accessible in GaN-based 2DEGs, here through the use of pulsed magnetic fields.

\begin{figure}[tbp]
\centering
\includegraphics[width=0.99\columnwidth]{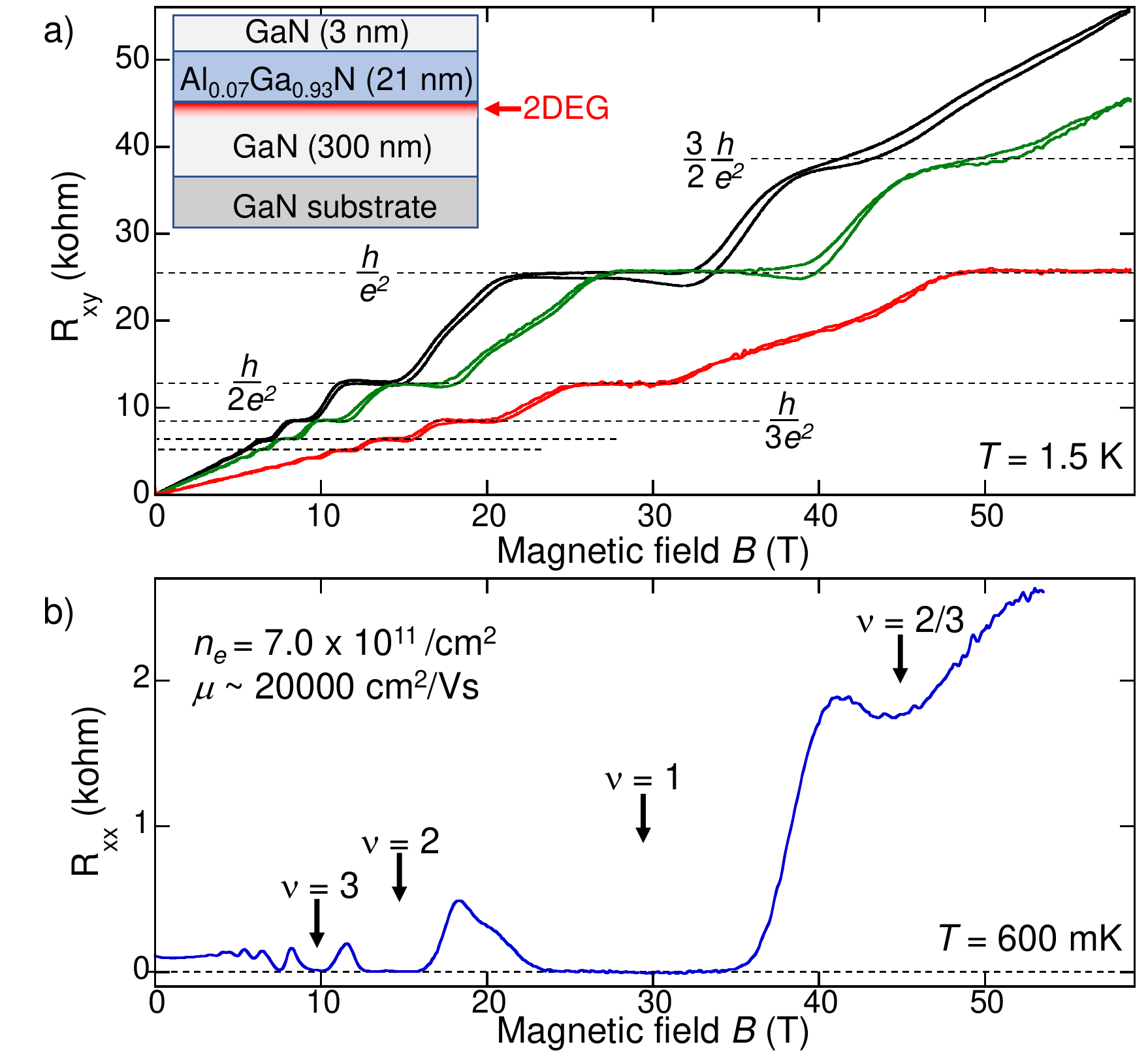}
\caption{a) Inset: schematic of the GaN/AlGaN 2DEG structure. The plot shows the transverse resistance ($R_{xy}$) at 1.5~K up to 60~T, showing quantized Hall resistance $h/\nu e^2$ at integer filling factors $\nu$. The black trace was acquired after three days in the dark, and the quantum limit ($\nu=1$) is reached at $\approx$27~T. The red trace was then acquired after briefly illuminating the structure \textit{in-situ} with white light -- note that $n_e$ approximately doubled. The green trace was then acquired after cycling the temperature up to 300~K and back to 1.5~K in the dark.  b) Longitudinal resistance $R_{xx}$ of this structure (separate cooldown; $\nu =1$ is reached at $\approx$29~T). Both $R_{xx}$ and $R_{xy}$ show evidence for the $\nu = 2/3$ FQHE state.}
\end{figure}

We now turn to the other $R_{xy}$ curves in Fig. 1a. After measuring the black curve, the sample was briefly and weakly illuminated at 1.5~K by an \textit{in-situ} white-light LED.  Then the red curve was measured in the dark. IQHE plateaus were again observed, but significantly shifted in field: $\nu=1$ occurred at $\approx$57~T, indicating over a two-fold increase of $n_e$. Then, the sample was thermally cycled (in the dark) up to 300~K for 6 hours and back to 1.5~K, and the green curve was measured. IQHE plateaus were again observed, but $\nu=1$ occurred at $\approx$32~T, indicating that $n_e$ had (mostly) recovered back down to its initial value. IQHE plateaus in all curves shows no obvious signs of inhomogeneity or disorder changes resulting from illumination.

As described in early \cite{Elhamri2000} and more recent \cite{Kruckeberg2020} studies, illumination has a significant and persistent effect on $n_e$ in GaN/AlGaN 2DEGs. Optical excitation of electrons out of traps and impurity states leads to a persistent increase of $n_e$, that can be reset by thermal cycling in the dark. These data point to the fact that -- at least in this particular structure -- FQHE states with $\nu < 1$ are not accessible with 60~T fields if the sample has been recently illuminated. This limits the use of optical spectroscopy to study FQHE states (in this structure), which in GaAs-based 2DEGs have historically proven very powerful for probing electron correlations and novel phases \cite{Goldberg1992, Nurmikko1993, Kukushkin1996, Heiman1988, Katayama1989, Goldberg1990, Buhmann1990, Turberfield1990, Byszewski2006}.   

Optical phenomena related to the IQHE can, however, be studied quite well in GaN/AlGaN 2DEGs, even in low-field ($<$20~T) superconducting magnets. The PL spectrum from this GaN/AlGaN structure is shown in Fig. 2a.  The two bright peaks at 3.481~eV and 3.474~eV correspond to free ``A''-excitons and to donor-bound excitons in the GaN buffer layer, respectively \cite{Kornitzer1999}. At lower energy, the broad and weak emission band in the $3.44 - 3.46$~eV range corresponds to radiative recombination of photogenerated holes with electrons in the 2DEG, as depicted in the inset and as confirmed  below. 

\begin{figure}[tbp]
\centering
\includegraphics[width=.99\columnwidth]{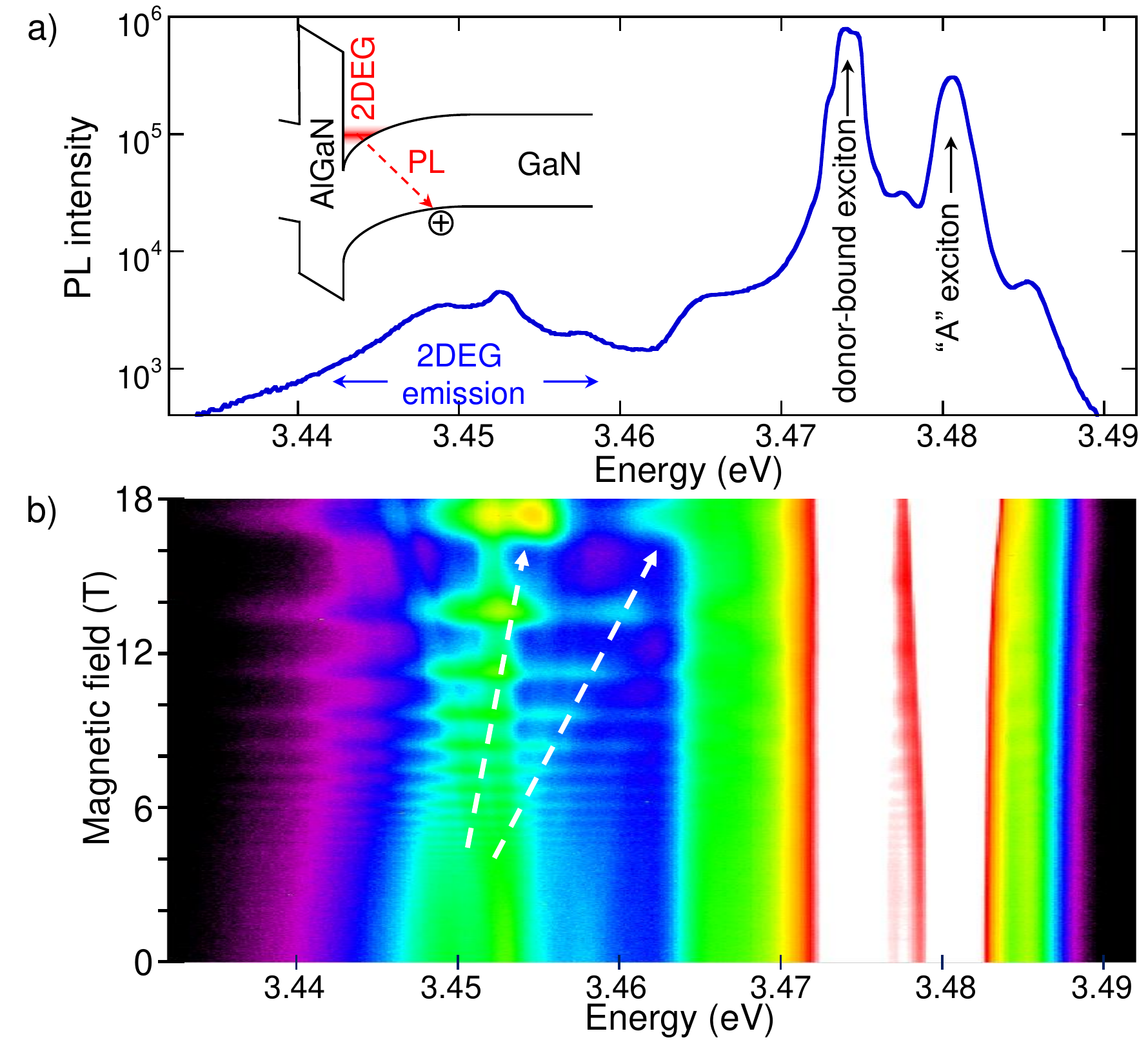}
\caption{a) Photoluminescence (PL) spectrum of the sample at 1.5~K; note logarithmic scale.  A broad PL emission from radiative recombination of 2DEG electrons is located at low energies ($\sim$3.45 eV), below the sharp exciton-related PL peaks that originate from the GaN buffer layer. b) An intensity map showing the PL spectra up to 18~T. SdH-like intensity oscillations are observed in the low-energy 2DEG region; discrete peaks reveal the formation and dispersion of 2DEG LLs (see white dashed arrows), separated by the cyclotron energy $eB/m_e$.}
\end{figure}

The PL intensity map of Fig. 2b shows how this low-energy emission band evolves with $B$ up to 18~T. Pronounced oscillations of the net PL intensity are observed, along with the formation of discrete peaks that shift linearly with $B$. These peaks reveal the formation and dispersion of discrete Landau levels in the 2DEG, separated by the electron cyclotron energy $eB/m_e$, from which an effective electron mass $m_e \simeq 0.24 m_0$ can be inferred. These data corroborate and extend recent results from Schmult \textit{et al.} \cite{Schmult2019}, who measured PL from a similar GaN/AlGaN 2DEG up to 15~T. 

Figure 3a shows how the PL intensity oscillations (red trace) compare with the simultaneously-measured longitudinal resistivity $R_{xx}$ (blue trace). Figure 3b shows the same data plotted against $1/B$: both curves have the same periodicity in $1/B$, as expected for Shubnikov-de Haas (SdH) quantum oscillations.  However, the minima in $R_{xx}$ -- which accurately indicate integer filling factors -- align only approximately with the maxima in the PL intensity.  The latter exhibit a relative phase shift that is most noticeable at large $B$. Moreover, intensity oscillations at odd-integer $\nu$, which manifest clearly in $R_{xx}$ for $B>7$~T, are not observed in the PL data up to 18~T. 

\begin{figure}[tbp]
\centering
\includegraphics[width=0.95\columnwidth]{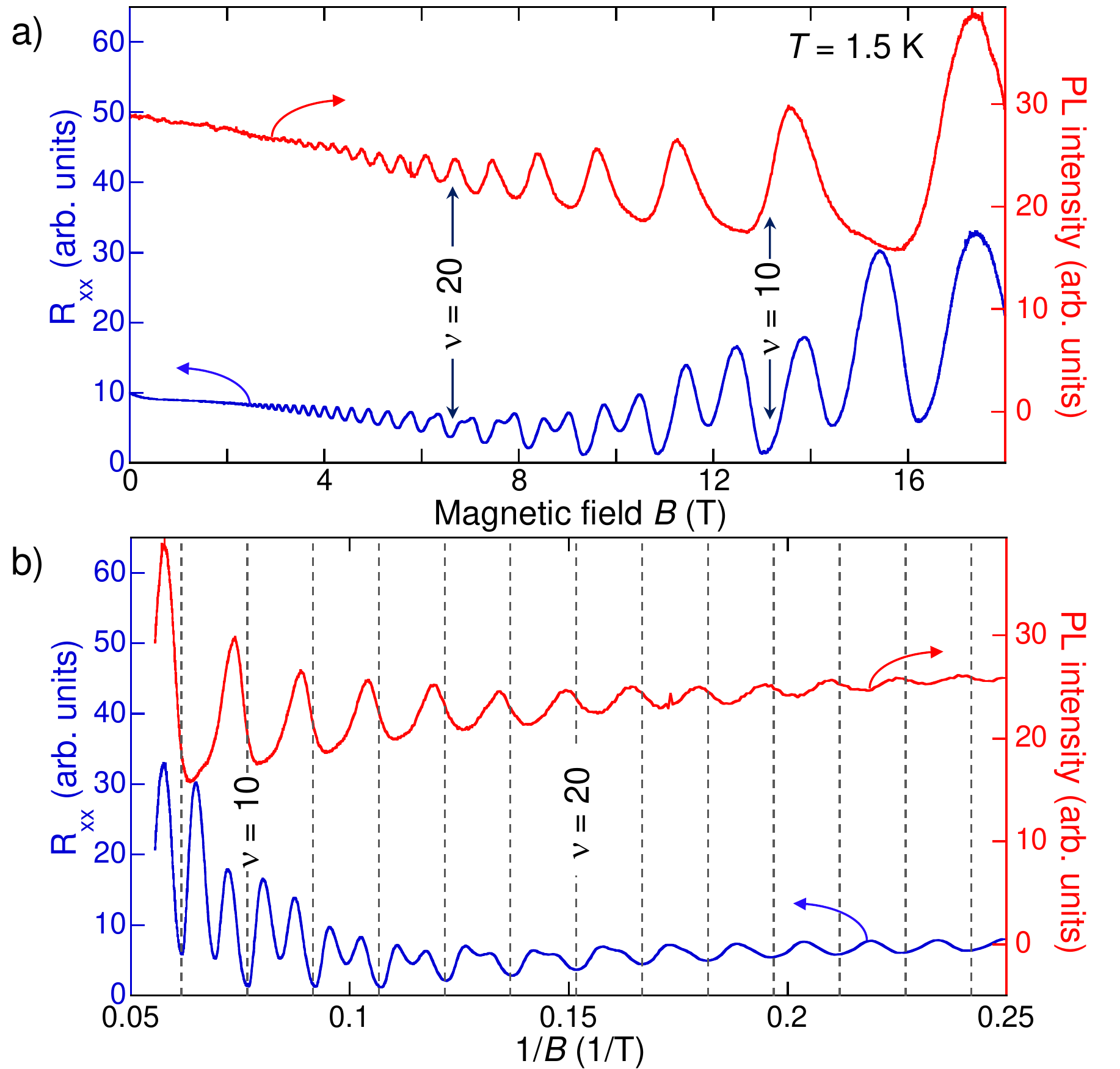}
\caption{a) Correlating the $B$-dependent oscillations of the total 2DEG PL intensity (integrated from $3.440 - 3.464$~eV; red trace) with the simultaneously-measured resistance $R_{xx}$ (blue trace) up to 18~T in a superconducting magnet, at 1.5~K. Due to the excitation laser, the steady-state 2DEG density is much higher than in the dark (\textit{cf.} Fig. 1); $n_e \simeq 3.1 \times 10^{12}$/cm$^2$, and $\nu=1$ is not expected until $\sim$130~T. b) The same data vs. $1/B$. Dashed vertical lines are equally spaced and aligned with $R_{xx}$ minima, and indicate even-numbered $\nu$. Both the PL intensity and $R_{xx}$ are periodic in $1/B$, but only the latter shows odd filling factors. Note that PL maxima are slightly shifted from even-numbered $\nu$.}
\end{figure}

Studies of SdH-like PL oscillations in 2DEGs have a long and rich history \cite{Goldberg1992, Nurmikko1993, Kukushkin1996, Kukushkin1988, Chen1992, Babinski2006, Solovyev2015}. Typically, they arise from electron-hole correlations and the efficacy with which 2D electrons screen the Coulomb potential of a nearby photogenerated hole, which in turn modifies the spatial overlap of their respective wavefunctions and therefore changes the radiative recombination rate. In many models, screening is less effective when the 2DEG Fermi level lies between Landau levels (\textit{i.e.}, in localized states, at integer $\nu$). On the one hand this can lead to a reduced wavefunction overlap and smaller PL intensity; however, reduced screening can also increase electron-hole binding (exciton formation), increasing PL \cite{Goldberg1992, Kukushkin1996}. The phase shift that is observed between the quantum oscillations of $R_{xx}$ and PL (see Fig. 3) -- which increases at large $B$ -- suggests that the interplay between these competing effects likely plays an important (and $B$-dependent) role.  Additionally, the fact that $R_{xx}$ clearly exhibits spin-resolved SdH oscillations (odd-integer $\nu$) for $B>7$~T, while the PL does not, also presents an unresolved puzzle (although, $\nu=3$ appears in PL studies to 60~T, shown below).

Importantly, Fig. 3 also reveals that due to the weak ($\sim$100~$\mu$W) above-gap optical excitation that is used to enable PL measurements, the steady-state electron density is about $5\times$ larger than when measured in the dark (\textit{cf.} Fig. 1). In Fig. 3, $\nu=10$ occurs at  13~T, indicating that $n_e \simeq 3.1 \times 10^{12}$/cm$^2$. A consequence of this photo-doping effect is that future optical studies of GaN-based 2DEGs in the $\nu \leq 1$ regime will require structures specifically designed to host lower carrier densities.  

Nonetheless, to explore the limits of high-$B$ magneto-optics in GaN-based 2DEGs, Fig. 4 shows PL studies of this same structure in a 60~T pulsed magnet. Due to the fast CCD exposure times that are used, fewer photons are collected, limiting signal:noise. Regardless, Fig. 4 clearly shows that the 2DEG emission continues to exhibit pronounced intensity oscillations. A maximum is observed at 17.5~T (the same as observed in Figs. 2 and 3), and then also at $B \approx$ 24~T, 39~T, and 46~T. With the exception of the 39~T peak, these values are in reasonable agreement with the expected filling factors $\nu$=8, 6, 4, and also $\nu$=3, as indicated.  The appearance of a PL maximum at the expected position of $\nu =3$ suggests that detailed optical studies of \textit{spin-resolved} many-body screening in GaN-based 2DEGs are possible. Future measurements incorporating in-situ polarizers to resolve both right- and left-handed circular PL polarizations, ideally in conjunction with transport measurements performed on intrinsically lower-density 2DEGs, are expected to address this interesting regime. 

\begin{figure}[tbp]
\centering
\includegraphics[width=0.95\columnwidth]{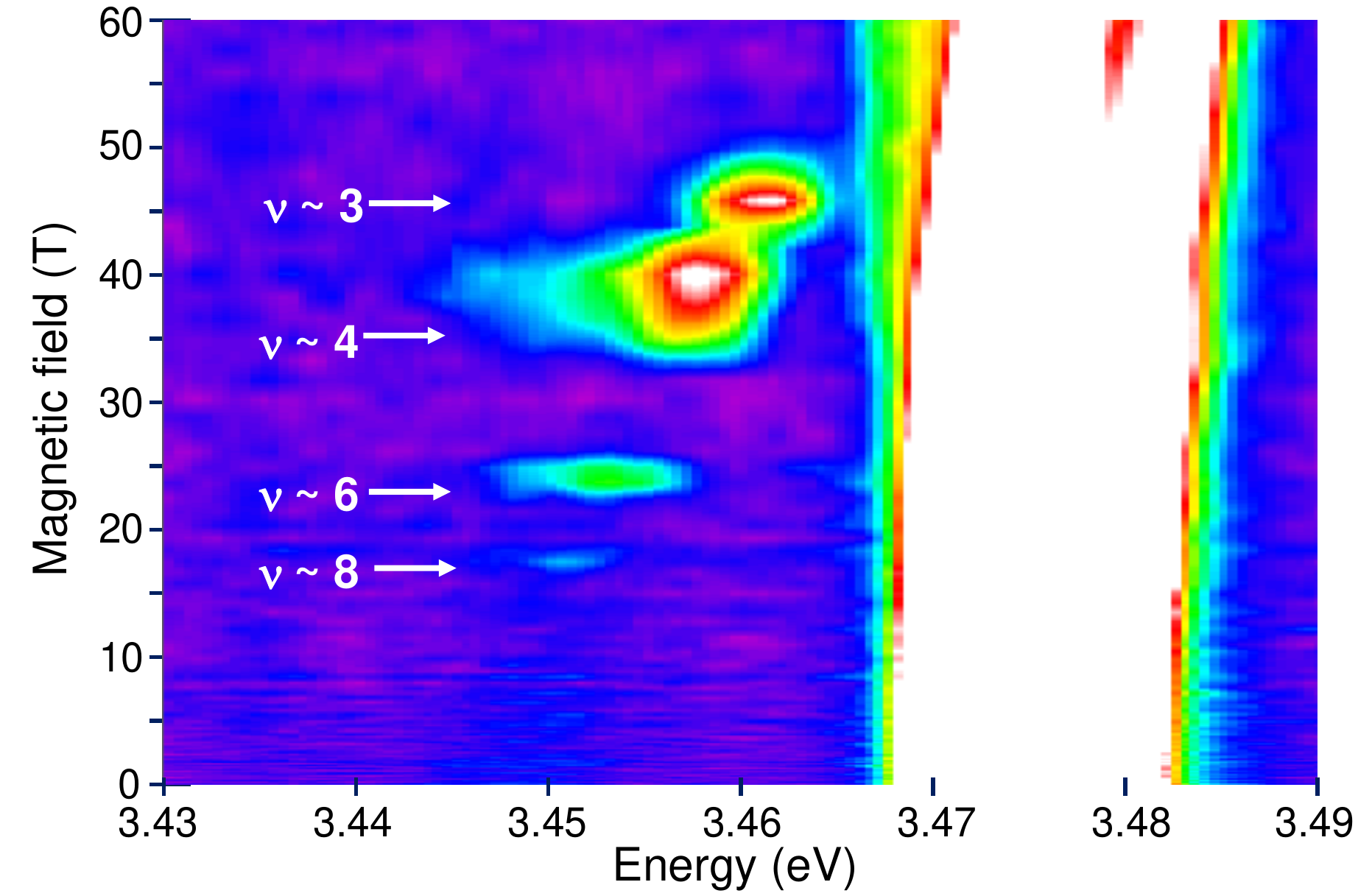}
\caption{PL measured up to 60~T. PL from the 2DEG brightens considerably at 17.5~T (as also seen in Figs. 2 and 3), 24~T, 39~T, and 46~T, which corresponds approximately to $\nu \approx$ 8, 6, 4, and 3. Limited signal:noise precludes resolving the smaller intensity oscillations for $B<$15~T.}
\end{figure}

Through the use of pulsed magnetic fields, these studies demonstrate that the $\nu=1$ quantum limit can be exceeded in high-density GaN-based 2DEGs. First evidence for the $\nu=2/3$ FQHE is shown by magneto-transport.  PL measurements also reveal SdH-like oscillations of the intensity, from which $n_e$ and the electron mass are inferred in a contactless approach.  However, the increase of $n_e$  associated with optical excitation currently limits PL studies to the IQHE regime, although strong indications of 2DEG screening at the spin-resolved $\nu = 3$ state is observed at large fields. 

Work at the National High Magnetic Field Laboratory was supported by National Science Foundation (NSF) DMR-1644779, the State of Florida, and the U.S. Department of Energy (DOE). Work at Cornell was supported in part by the Office of Naval Research (grants N00014-20-1-2126 and N00014-20-1-2176) monitored by Dr. Paul Maki, the NSF RAISE TAQs Award No. 1839196 monitored by Dr. D. Dagenais, and the NSF CCMR MRSEC Award No. 1719875. S.A.C. acknowledges support from the DOE BES `Science of 100~T' program. B.J.R. acknowledges funding from NSF DMR-1752784. 



\begin{thebibliography}{10}

\bibitem{Ambacher1999} O. Ambacher, J. Smart, J. R. Shealy, N. G. Weimann, K. Chu, M. Murphy, W. J. Schaff, and L. F. Eastman, Two-dimensional electron gases induced by spontaneous and piezoelectric polarization charges in N- and Ga-face AlGaN/GaN heterostructures, J. Appl. Phys. \textbf{85}, 3222 (1999).

\bibitem{Frayssinet2000} E. Frayssinet \textit{et al.}, High electron mobility in AlGaN/GaN heterostructures grown on bulk GaN substrates, Appl. Phys. Lett. \textbf{77}, 2551 (2000).

\bibitem{Elhamri2000} S. Elhamri \textit{et al.}, Persistent photoconductivity study in a high mobility AlGaN/GaN heterostructure, J. Appl. Phys. \textbf{88}, 6583 (2000).

\bibitem{Manfra2004} M. Manfra, K. W. Baldwin, A. M. Sergent, K. W. West, R. J. Molnar, J. Caissie, Electron mobility exceeding 160000 cm$^2$/Vs in AlGaN/GaN heterostructures grown by molecular-beam epitaxy, Appl. Phys. Lett. \textbf{85}, 5394 (2004). %

\bibitem{Skier2005} C. Skierbiszewski \textit{et al.}, High mobility two-dimensional electron gas in AlGaN/GaN heterostructures grown on bulk GaN by plasma assisted molecular beam epitaxy, Appl. Phys. Lett. \textbf{86}, 102106 (2005).

\bibitem{Knap2004} W. Knap \textit{et al.}, Spin and interaction effects in Shubnikov-de Haas oscillations and the quantum Hall effect in GaN/AlGaN heterostructures, J. Phys. Cond. Matt. \textbf{16}, 3421 (2004). %

\bibitem{Suzuki2018} K. Suzuki and T. Akasaka, Landau level quantization with gate tuning in an AlN/GaN single heterojunction, Japanese J. Appl. Phys. \textbf{57}, 111001 (2018). %

\bibitem{Schmult2019} S. Schmult, V. V. Solovyev, S. Wirth,  T. Mikolajick, and I. V. Kukushkin, Magneto-optical confirmation of Landau level splitting in a GaN/AlGaN 2DEG grown on bulk GaN, J. Vac. Sci. Tech. B \textbf{37}, 021210 (2019).  

\bibitem{Kruckeberg2020} L. Kr\"{u}ckeberg, S. Wirth, V. Solovyev, A. Grosser, I. V. Kukushkin, T. Mikolajick, S. Schmult, Quantum and transport lifetimes in optically induced GaN/AlGaN 2DEGs grown on bulk GaN, J. Vac. Sci. Tech. B \textbf{38}, 042203 (2020).

\bibitem{Manfra2002} M. Manfra \textit{et al.}, High mobility AlGaN/GaN heterostructures grown by plasma-assisted molecular beam epitaxy on semi-insulating GaN templates prepared by hydride vapor phase epitaxy, J. Appl. Phys. \textbf{92}, 338 (2002). 

\bibitem{Falson2018}J. Falson and M. Kawasaki, A review of the quantum Hall effects in MgZnO/ZnO heterostructures, Rep. Prog. Phys. \textbf{81}, 056501 (2018)

\bibitem{ManfraMobility}M. J. Manfra, K. Baldwin, A. M. Sergent, R. J. Molnar, J. Caissie, Electron mobility in very low density GaN/AlGaAs/GaN heterostructures, Appl. Phys. Lett. \textbf{85}, 1722 (2004).

\bibitem{Cho2019}Y. Cho, Y. Ren, H. G. Xing, and D. Jena, High-mobility two-dimensional electron gases at AlGaN/GaN heterostructures grown on GaN bulk wafers and GaN template substrates, Appl. Phys. Exp. \textbf{12}, 121003 (2019).

\bibitem{Battesti2018} R. Battesti \textit{et al.}, High magnetic fields for fundamental physics, Physics Reports \textbf{765-766}, 1 (2018).

\bibitem{Goldberg1992} B. B. Goldberg, D. Heiman, A. Pinczuk, L. Pfeiffer, K. W. West, Magneto-optics in the integer and fractional quantum Hall and electron solid regimes, Surf. Sci. \textbf{263}, 9 (1992).

\bibitem{Nurmikko1993} A. Nurmikko and A. Pinczuk, Optical probes in the quantum Hall regime, Physics Today \textbf{46}, 24 (1993).

\bibitem{Kukushkin1996} I. V. Kukushkin and V. B. Timofeev, Magneto-optics of strongly correlated two-dimensional electrons in single heterojunctions, Adv. Phys. \textbf{45}, 147 (1996). %

\bibitem{Heiman1988} D. Heiman, B. B. Goldberg, A. Pinczuk, C. W. Tu, A. C. Gossard, and J. H. English, Optical Anomalies of the Two-Dimensional Electron Gas in the Extreme Magnetic Quantum Limit, Phys. Rev. Lett. \textbf{61}, 605 (1988).

\bibitem{Katayama1989} S. Katayama and T. Ando, Magnetic oscillation of luminescence energy in modulation-doped quantum wells, Solid State Comm. \textbf{70}, 97 (1989).

\bibitem{Goldberg1990} B. B. Goldberg, D. Heiman, A. Pinczuk, L. Pfeiffer, and K. West, Optical investigations of the integer and fractional quantum Hall effects: Energy plateaus, intensity minima, and line splitting in band-gap emission, Phys. Rev. Lett. \textbf{65}, 641 (1990).

\bibitem{Buhmann1990}H. Buhmann, W. Joss, K. von Klitzing, I. V. Kukushkin, G. Martinez, A. S. Plaut, K. Ploog, and V. B. Timofeev, Magneto-optical evidence for fractional quantum Hall states down to filling factor 1/9, Phys. Rev. Lett. \textbf{65}, 1056 (1990).

\bibitem{Turberfield1990} A. J. Turberfield, S. R. Haynes, P. A. Wright, R. A. Ford, R. G. Clark, J. F. Ryan, J. J. Harris, and C. T. Foxon, Optical detection of the integer and fractional quantum Hall effects in GaAs, Phys. Rev. Lett. \textbf{65}, 637 (1990).

\bibitem{Byszewski2006}M. Byszewski \textit{et al.}, Optical probing of composite fermions in a two-dimensional electron gas, Nat. Phys. \textbf{2}, 239 (2006).

\bibitem{Solovyev2015} V. V. Solovyev \textit{et al.}, Optical probing of MgZnO/ZnO heterointerface confinement potential energy levels, Appl. Phys. Lett. \textbf{106}, 082102 (2015).

\bibitem{Goryca2019} M. Goryca \textit{et al.}, Revealing exciton masses and dielectric properties of monolayer semiconductors with high magnetic fields, Nat. Commun. \textbf{10}, 4172 (2019).

\bibitem{Kornitzer1999} K. Kornitzer, T. Ebner, K. Thonke, R. Sauer, C. Kirchner, V. Schwegler, M. Kamp, M. Leszczynski, I. Grzegory, and S. Porowski, Photoluminescence and reflectance spectroscopy of excitonic transitions in high-quality homoepitaxial GaN films, Phys. Rev. B \textbf{60}, 1471 (1999).

\bibitem{Kukushkin1988} I. V. Kukushkin, K. von Klitzing, K. Ploog, Optical spectroscopy of two-dimensional electrons in GaAs-Al$_x$Ga$_{1-x}$As single heterojunctions, Phys. Rev. B \textbf{37}, 8509 (1988).

\bibitem{Chen1992} W. Chen, M. Fritze, W. Walecki, A. V. Nurmikko, D. Ackley, J. M. Hong, and L. L. Chang, Excitonic enhancement of the Fermi-edge singularity in a dense two dimensional electron gas, Phys. Rev. B \textbf{45}, 8464 (1992). %

\bibitem{Babinski2006} A. Babi\'{n}ski, M. Potemski, and H. Shtrikman, Quantum oscillations of the luminescence from a modulation-doped GaAs/InGaAs/GaAlAs quantum well, Appl. Phys. Lett. \textbf{88}, 051909 (2006).

\end{thebibliography}
\end{document}